# AN EXAMINATION OF THE EFFECTIVENESS OF TEACHING DATA MODELLING CONCEPTS

Márta Czenky

Department of Informatics, Szent István University, Gödöllő, Hungary

## ABSTRACT

*The effective teaching of data modelling concepts is very important; it constitutes the fundament of database planning methods and the handling of databases with the help of database management languages, typically SQL. We examined three courses. The students of two courses prepared for the exam by solving tests, while the students of the third course prepared by solving tasks from a printed exercise book. The number of task for the second course was 2.5 times more than the number of task for the first course. The main purpose of our examination was to determine the effectiveness of the teaching of data modelling concepts, and to decide if there is a significant difference between the results of the three courses. According to our examination, with increasing the number of test tasks and with the use of exercise book, the results became significantly better.*

## KEYWORDS

*Data modelling, Teaching of Data Modelling Concepts, Survey, Statistical Analysis, Data Mining Analysis*

## 1. INTRODUCTION

At the Department of Informatics of Szent István University we teach database management for some faculties and majors. We selected three courses where the curriculum of the subjects did not differ, but the learning circumstances were different.

The two main topics of the Database Management subject attended by mechanical engineer students at Faculty of Mechanical Engineering are data modelling and the SQL-92 language. The duration of the subject is one semester with two classes per week. Classes on data modelling take half of the semester. We supported the learning of definitions and properties of concepts and the effective identification and realization of the concepts with tests generated from the question bank of the Moodle e-learning education system. We included two courses in the analysis we are going to reference these as ABK1 and ABK2. For the first course there were 92 tasks in Moodle question bank, for the second course this number was 250. At the exam the students had to solve tests.

At Faculty of Agricultural and Environmental Sciences the subject of Computer Studies III is taught for environment engineering students. The duration of the subject is half a semester, including two classes of lectures and two classes of seminars per week. In the first half of the semester the students learn data modelling at the lectures and SQL-92 language at the seminars. In the second part of the semester they learn CAD knowledge. A printed data modelling exercise book helped their preparation, but they had to solve a test at the exam [1]. Beside the exercise book, a practice test was also available for students, consisting of 25 randomly selected tasks, so they can try this kind of exam. We refer to this course as KM3 in the article.

We summarized in Table 1 the characteristics of the subjects. All students of the courses participated in the analysis therefore we can consider the results representative.

Table 1. The characteristics of subjects

| Name of subject | Abbre-viation | Major | Level/ Course | Class per week lecture+ practise | Head-count |
|---|---|---|---|---|---|
| Database Management | ABK1 ABK2 | mechanical engineer | BSc, full time | 0+2 | 64 19 |
| Computer Studies III. | KM3 | environmental engineer | BSc, full time | 2+2 | 71 |

We are going to compare the results of electronic tests for the three courses. Based on this, we are going to decide whether our teaching methods are efficient enough, and if there is a significant difference between the results of the courses.

Our questions are the following:

- Which areas of concept learning can be considered good enough? We regard those areas sufficient where the students achieved satisfactory or better result.

- In which topics and task types are the results below average?

- Did the increase in the numbers of tasks and tests improve the results significantly?

- Is there a significant difference between the effectiveness of different learning methods, such as computational tests or printed exercise book?

- Can we determine the expected score of the student by task type and topic?

## 2. TEACHING OF CONCEPTS

The concept is a mental form developed from the main characteristics of the things in mind [2], [3]. In our course at the constitution of the concept we disregard the irrelevant features and we unite the essential characteristics into a standardized form. The content of the concept is the totality of the substantive characteristics. The extent of the concept is given by the things, which dispose the substantive characteristics creating the concept.

The main topics of data modelling are: basic data modelling concepts, the entity relationship model, the relational model, dependencies, dependence diagrams and normalization. Based on these we can classify the concepts of the data modelling into the following groups:

- The basic concepts of data modelling are: entity and set of entities, relationship and occurrence of relationship, attribute and value of attribute. These concepts also appear at data models, but they are not listed separately there.

- Entity relationship model (E-R): hierarchy of set of entities, ISA relationship, inheritance of attributes, integrity constraints.

- Relational model: relational table, keys, special data values, integrity constraints and their enforcements, objects of relational database (index, view, synonym), relational operations.

- Dependences: functional and multivalued dependence.

- Dependence diagrams: the depiction of relational dependences of a table on diagram.

- Normalization: normal forms [4], [5], [6].

The concepts may be objective concepts, for example entity, set of entities, relational table, primary key, etc., relational concepts, for example relationship, occurrence of relationship, etc. and operational concepts, for example relational operations, cascade update and deletion, etc.

The order of the teaching of topics may differ; implicit teaching of them begins with the teaching of the basic data modelling concepts. Afterwards you may teach entity relationship model or relational model, finally dependences and normalization. [7] follows the first way while [8] the second one. Currently we teach the relational model first, following this the entity relationship model and its transcription into relational model. Based on our experience the entity relationship model is more understandable for students than relational model. If we begin teaching with E-R model though, then at one point we have to stop with it, and we can only introduce the transcription of the E-R model into relational model after teaching relational model.

There are more ways to introduce the concepts:

- the inductive way, that is to get from the concrete to the abstract,

- the deductive way, that is to get from abstract to the concrete,

- the constructive way, to produce a representative, then the generalization of the procedure.

While teaching the concepts we primarily follow the deductive way, but the discussion of relational model happens inductively. In the database system we introduce the tables filled up with data, their structure and relationships, after this we define the essential characteristics of the concepts of relational model. [9] describes constructive way of teaching of the relational model.

The efficient concept learning means the acquirement of the following knowledge and abilities:

- the knowledge of the definition of the concept,

- concept identification, that is, you can decide whether a thing belongs to a certain concept or not,

- concept realization, that is the recognition and enumeration examples,

- knowledge of the attributes of the concept,

- concept systematization, that is the knowledge of the relationship between concepts,

- concept classification, that is placement the concept into the hierarchical system of the concepts; the concept hierarchy shows the subordinate, the superior or the coordinate relationship with each other,

- the usage of the concept for the description of situations, and for the solution of problems [10], [11], [12], [13].

[14] emphasizes to the importance of the concept learning in database planning. He emphasizes the importance of correct concept identification and the knowledge of the attributes of the concepts, which are necessary for the categorization of basic concepts. [15] also emphasizes the importance of these two steps of concept learning in connection with the teaching of the E-R model.

The tasks of the Moodle question bank and exercise book used by us support the listed steps of concept learning. The main part of tasks is smaller modelling task with which the students can practise problem solving.

We designed separate sheets for the execution of the concept systematization and classification. At first, students had to fill in Venn diagrams and hierarchical figures of them but lately we changed it to concept maps of data modelling, drawn by us, which reflect the relationships between concepts more complex than the concept hierarchies [13].

## 3. THE STUDENTS' JUDGEMENT OF CONCEPT LEARNING

A survey with 62 questions was made among the students of ABK1 course at the end of the semester to get to know the opinion of students about education of database management, including their view about concept learning. We designed the questionnaire by instruction of [16], 54 stu-dents filled it. It was random who did not fill in the survey so the survey can be considered representative.

In the next three tables we present the answers given for the questions related to data modelling.

In the questionnaire we asked the students what kind of activities meant trouble for them in the course of the class work? The answers are summarized in Table 2. According to the students' opinion the most problematic activity is the comprehension of algorithms, followed by the understanding of concepts and definitions. It causes fewer problems to find the examples, but this contradicts our teacher experiences.

Table 2. Problematic activities during class work

| Activity | Proportion of students indicating problems |
|---|---|
| usage of terminology | 5.54% |
| understanding of the concepts and the definitions | 20.36% |
| understanding of the example | 12.99% |
| finding of the example | 7.41% |
| comprehension of the algorithm | 22.21% |
| application of the algorithm | 16.63% |

We indicated in Table 3 the percentage of students having problems in understanding different concepts of data modelling .The most problematic concepts are normalization, the normal forms and the different dependences. From the basic data modelling concepts relationship and occurrence of relationship are mentioned as problematic, from the relational model concepts the foreign key and the integrity constraints seemed to cause a problem.

Table 3. Proportion of students indicating problems with the comprehension of concepts

| Concept | Proportion of students indicating problems |
|---|---|
| entity, set of entities | 5.51% |
| occurrence of relationship, relationship | 20.38% |
| attribute, value of attribute | 5.54% |
| primary key | 7.39% |
| foreign key | 14.84% |
| Null value | 1.85% |
| functional dependence | 44.44% |
| multivalued dependence | 42.63% |
| normal forms | 68.55% |
| integrity constraints | 22.25% |

The activities listed in the Table 4 belong to the area of concept identification and concept realization. Students were asked how problematic they had found the recognition of the most important concepts of the data modelling: that is, the recognition of the basic modelling concepts and the different dependences. The answers show that recognition of the dependences was much more problematic than recognition of basic modelling concepts. Although students do not indicate major problems with recognition of relationships, our teacher experience is that this is much more complicated for them, than the recognition of the sets of entities and attributes [17].

Table 4. The proportion of students indicating problems with the recognition of the concepts

| Activity | Proportion of students indicating problems |
|---|---|
| recognition of the sets of the entities | 5.51% |
| recognition of the relationships | 9.26% |
| recognition of the attributes | 7.39% |
| recognition of the functional dependences | 29.66% |
| recognition of the multivalued dependences | 24.10% |

We examined too if there is any difference between the answers of the students who achieved better or worse result. We found that the students, who achieved better results, considered the more easily understandable concepts less problematic but found difficult concepts problematic in a much higher proportion than those students who received worse mark. The explanation to this could be that students achieved better marks understood and learned the concepts better, and at the end of the semester they were more likely to know which concept the respective questions applied to.

## 4. E-LEARNING AND EXERCISE BOOK IN THE CONCEPT TEACHING

The learning of the concepts was supported by tests created from tasks of the Moodle course management system (ABK courses), and the tasks of the printed data modelling exercise book (KM3 course). The measurement of concept comprehension was done through electronic exam tests. The exam tests of ABK1 and KM3 courses were randomly generated from the question bank so every student solved different test. The exam tests of ABK2 course consisted of the same questions for all, so every student solved the same tasks. Although the exam tests were diverse, their tasks were from the same topic and they included the same type of task.

The practising and exam tests of the ABK1 course were different from the similar test of ABK2 course. For the ABK2 course we increased the number of tasks in the question bank, as well as the number of the practicing and exam tests. The students of the KM3 course prepared for the exam with the help of a printed data modelling exercise book. The exercise book includes the tasks of the question bank of ABK2 course, and also included some new tasks. The exam test of KM3 course was different from the exam tests of the first two courses. We made a practise test for KM3 course so that the students could try this exam method.

Table 5. The number of tasks of question bank/exercise book and tests

| Course | Preparation method | Number of tasks in question bank/exercise book | Number of practice tests | Number of exam tests |
|---|---|---|---|---|
| ABK1 | test | 92 | 5 | 1 |
| ABK2 | test | 250 | 14 | 6 |
| KM3 | exercise book | 300 | 1 | 1 |

We summarized in Table 5 the preparation methods of the groups, the number of the tasks in the question bank/exercise book and the number of practising and exam test.

The tasks of the tests and the exercise book require the knowledge of definitions, concept identification and knowledge of concept characteristics. At concept realization, the listing examples can only be evaluated manually by the teacher. Therefore we always asked the enumeration of the examples at the class. In the tests we coupled those tasks to the concept realization where examples for a concept had to be recognized.

Table 6. Abbreviations used in figures

| Topic | Abbre-viation | Task | Abbre-viation |
|---|---|---|---|
| Basic modelling concepts | BAS | giving definitions | DEF |
| E-R model | ERM | knowledge of the concept characteristics | CHAR |
| Relational model | REM | concept identification | IDEN |
| Dependences | DEP | concept realization | REAL |
| Normalization | NOR | concept collation | COL |
| Dependence diagrams | DEPD | mapping of different model's concepts | MAPM |
| | | mapping of concepts of different depiction | MAPD |

Furthermore, there are also tasks for the measurement of concept learning:

- In the course of learning basic modelling concepts, concept collation tasks includes tasks like identification of relationships between sets of entities and mapping of attributes to a set of entities or to a relationship. In the course of the learning of relational model concepts, collation tasks were to answer what kind of basic modelling concept (set of entity, relationship, both) is shown in the table.

- In E-R modelling at the mapping of different concepts of models the task was transcription of E-R diagrams into relational models.

- In the case of dependence diagrams at the mapping of the concepts of the different depiction the task was the mapping of different dependences described with dependence diagram and schema.

We classified all tasks in the Moodle question bank by topic and by type of concept learning, which allowed us to summarise and examine the results by topic and task type. In certain figures in this article we used abbreviations to spare space. These abbreviations are summarized in Table 6.

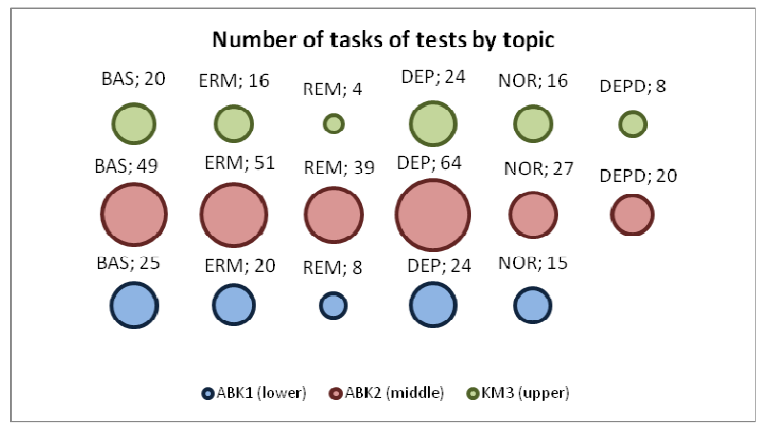

Figure 1. Distribution of the tasks by topics

Figures 1 and 2 show the distribution of the tasks of the electronic tests by topic and concept learning task type, respectively. Figure 1 shows that the four emphasized topics are: basic modelling concepts, relational model, dependences and normalization. These topics are the ones ensuring the necessary comprehension for the planning and management of relational databases. The E-R model and the dependence diagrams provide an alternative in database design.

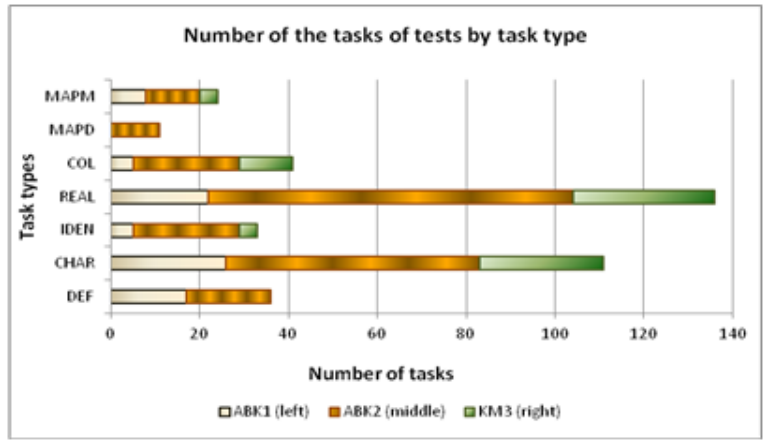

Figure 2. Distribution of the tasks by task types

The distribution of tasks of the exercise book by topic and concept learning task types is similar to the distribution of the tasks of the question bank of the ABK2 course.

According to Figure 2 most tasks are connected to concept realization. The explanation of this is that there are seven normalization tests among the tests of the ABK2 course, with 7-8 questions per test. Apart from tasks on the decomposition of tables, every task is a concept realization task. Data modelling concepts have numerous attributes, whose knowledge is necessary for database design. This explains the large number of tasks connected to the attributes of the data modelling concepts.

We took into consideration the results of answers given to tasks of the electronic tests. We did not ask students to evaluate the solutions of the tasks of the exercise book, so we do not have information about it. At each task the incorrect answer is worth 0 point, the correct answer 1 point. We did not apply penalty points, there was not negative point, but you could get fraction point for the partly correct answer.

## 5. INVESTIGATIONAL METHODS

First we summarized the scores achieved in test trials by topic and task type, then dividing this score with the number of trials, we determined the results. We show this data in a table and we visualized the result of all tests and exam tests separately. We determined the mean and standard deviation of data series and we highlighted the values below mean with bold and italic font style, indicating that the concept learning is problematic in these topics and task types.

We also summarized the scores by task, and we divided the sum with the trial number of the tasks. It was necessary due to the use of random questions, which means that different questions were included in the tests variable times. Afterwards we determined the means and standard deviations at every data series than we examined with Chi-square test if the data have normal distribution.

Between data series totalized by task we made homogeneity test by pairs to decide if the results have same distribution, or there is a significant difference between them. If the data series have normal distribution, than knowing that the data series are independent, we verified the equality of standard deviations with F-test. If the standard deviations were equal, then we examined the equality of means with T-test, if not, then we used Welch-test. In case of data series with not normal distribution, we examined the equality of standard deviations with Levene-test, while the equality of means with Mann-Whitney-test. This way we could decide if the results have the same distribution, or any of them is significantly better [18], [19], [20].

At last, from the data of the exam tests of all three courses, we tried to make decision trees with test scores achieved by test trial and by task. The decision trees show which scores can be expected by topic and task type, which can indicate the problematic areas of concept learning [21].

## 6. THE RESULTS OF THE EXAMINATION AND THEIR ASSESSMENT

We summarized in Table 7 the results by topic and by task type. In some columns of the table we represented the results of all tests (practical and exam), while in others the results of electronic exam tests only. In the column we indicated with bold and italic font style those values which are below the mean of values of the column.

Table 7. The results of the concept learning

| Topic | Task | Proportion of the good solution (%) | | | | | |
|---|---|---|---|---|---|---|---|
| | | **ABK1** | | **ABK2** | | **KM3** | |
| | | **All** | **Exam** | **All** | **Exam** | **All** | **Exam** |
| Basic modelling concepts | giving definition | 82.9 | 89 | 82.9 | | | |
| | knowledge of concept characteristics | 63.5 | 71.1 | ***64.7*** | ***66*** | ***50.3*** | 76.5 |
| | concept identification | 73.6 | 86.9 | 73.5 | 81 | ***54.3*** | 83.3 |
| | concept realization | ***52.1*** | ***65.6*** | 56.5 | 65.5 | 58.4 | 77.4 |
| | concept collation | 69.8 | 77.7 | 78.4 | ***64.3*** | 61.6 | 84.8 |
| E-R model | knowledge of concept characteristics | ***52*** | ***65.6*** | 76.7 | ***63.8*** | 56.6 | 76.1 |
| | mapping of the different model's concepts | ***55.2*** | 74.7 | ***68.5*** | 75.5 | ***47.1*** | ***55.1*** |
| Relational model | giving definition | 79.8 | 89.6 | 77.5 | | | |
| | knowledge of concept characteristics | 69.6 | 76.4 | 74.4 | 71.5 | ***54.5*** | 77.5 |
| | concept realization | | | ***50.1*** | ***32*** | | |
| | concept collation | | | ***53.6*** | 74 | ***47.7*** | ***52.1*** |
| Dependences | giving definition | 67.1 | 77.5 | 75.1 | | | |
| | knowledge of concept characteristics | ***24*** | ***33.3*** | 78.2 | | | |
| | concept identification | | | ***55.9*** | ***65*** | | |
| | concept realization | 60.6 | 78.4 | ***67.8*** | 72.5 | 73.75 | 83.3 |
| Normalization | giving definition | 60.3 | 76.3 | 72.8 | | | |
| | concept realization | ***39*** | ***57.4*** | ***70.6*** | ***62.3*** | ***46.9*** | 79.7 |
| Dependence diagrams | knowledge of concept characteristics | | | 80.3 | 77.7 | ***50.2*** | ***66.2*** |
| | concept realization | | | 83.3 | | 58.1 | 89 |
| | mapping of the concepts of the different depiction | | | 76.4 | 74.3 | | |
| | mean of the values of the column | 60.7 | 72.8 | 70.9 | 67.5 | 55.0 | 75.1 |

Some cells of the table are empty, the reasons are the followings:

- We first used the Moodle system when teaching the ABK1 course. The development of the tasks happened parallel with teaching. We did not have enough time to develop all tasks in every topic and task type by the end of the semester.

- In exam tests of the ABK2 and the KM3 courses and in the practise tests of the KM3 course there were no tasks referring to the knowledge of the definition of the concepts. The reason behind this is that at the tests of ABK1 course and practise tests of ABK2 course we experienced that students learned definitions well. In the exam tests we considered the solution of small modelling tasks more important.

- There was time limit for the solutions of the exam tests because the students had a SQL exam on the same class. Therefore at the ABK2 and the KM3 courses there were no questions for every topic and task type.

By the data of Table 7 we can state the followings:

- With increasing of number of tests and questions the results were better, taking into consideration the results of all tests – ABK1 all and ABK2 all data series.

- The results of the ABK2 exam tests are worse compared to the result of the ABK1 exam test. Searching for the reason we analysed how much time did the students spend with solving the practice tests. We experienced that students of ABK2 course solved practice tests in a shorter time. It seems that they were only eager to know the correct solutions of the tests shown after finishing the tests. This way their modelling facility developed less, which played a part in the result of exam tests [22].

- The result of the exam test of the students of the KM3 course who learned from exercise book was better than the results of the exam tests of the other two courses.

- The results of all tests on the KM3 course were the worst. Students of this course solved the only one practice test many times, but their results were extremely bad, and this influences the summarized result, too.

At which topics and task types did the students achieve bad result? Hereinafter we list those topics and task types where the result is below the mean at more courses:

- basic modelling concepts: knowledge of concept characteristics and concept realization,

- E-R model: knowledge of concept characteristics and mapping of the different model's concepts,

- relational model: concept realization and concept collation,

- dependences: knowledge of concept characteristics and concept identification,

- normalization: concept realization,

- dependence diagrams: knowledge of concept characteristics.

You can see that almost every topic and task type occur in the list. Therefore we examine further this question with data mining methods.

We summarized the result by topic and task type and depicted them on diagrams, see Figure 3 and 4.

According to Figure 3, the result was worst in entity-relationship model and normalization topics, while the best in basic modelling concept and dependence diagrams topics. By the opinion of the students normalization is also the most difficult topic. The entity-relationship modelling is easily understandable and can be taught easily based on our teaching experience. Therefore we expected better result in the solution of these tasks.

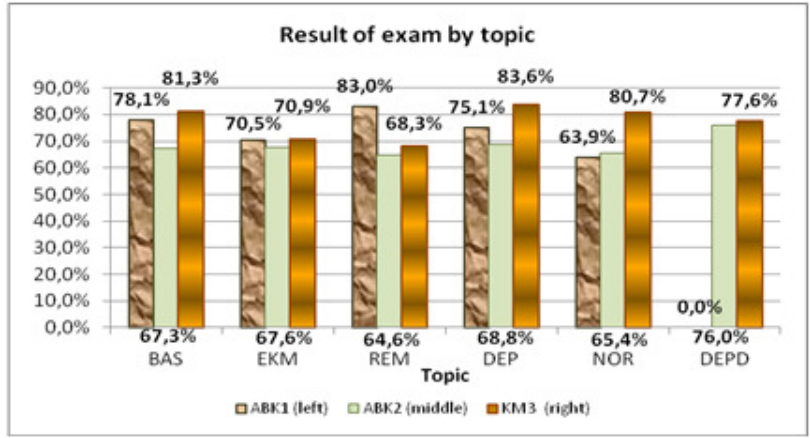

Figure 3. Summarized results by topics

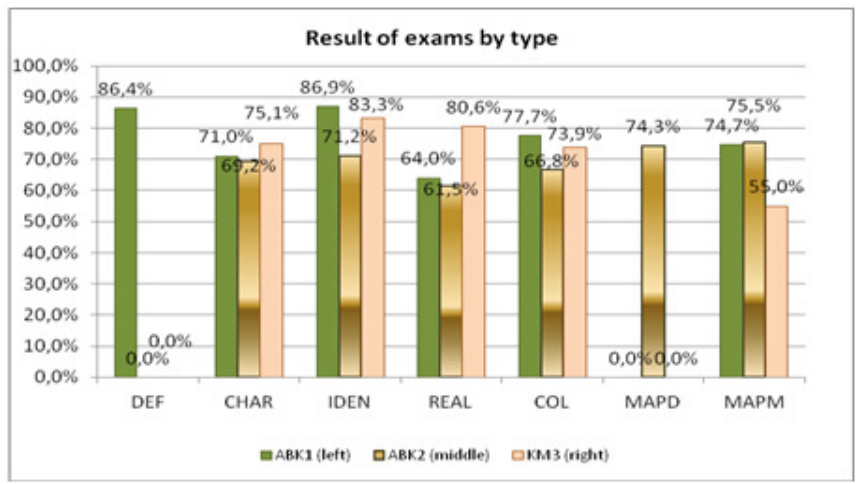

Figure 4. Summarized results by task type

Based on Figure 4 we can say that concept realization and mapping of the different model's concepts were the most problematic activities, while students can execute giving definition and concept identification the best.

To decide if the differences between results are significant, we executed homogeneity tests by the results, summarized by tasks between the following data series:

- ABK1 all – ABK2 all,
- ABK1 exam – ABK2 exam,
- ABK1 exam – KM3 exam,
- ABK2 exam – KM3 exam.

We executed the statistical examination in R program. In all cases the applied significance level was 0.05.

We made Chi-square test for all data series to decide if the data series have normal distribution. Our null hypothesis is:

*H0: The data series has normal distribution.*

We reject the null hypothesis if the probability value available after the execution of the test is smaller than significance level, keep it otherwise. Table 8 includes the results of the Chi-square tests.

In Table 8 we also show the means and the standard deviations by data series.

Table 8. Means and standard deviations

| Characteristic | Data series | | | | |
| --- | --- | --- | --- | --- | --- |
| | **ABK1 all** | **ABK2 all** | **ABK1 exam** | **ABK2 exam** | **KM3 exam** |
| **Chi-square test p/assessment** | 0.8554<br>normal | 0.00024<br>not normal | 0.1009<br>normal | 0.00375<br>not normal | 0.02123<br>not normal |
| **mean** | 0.6012 | 0.7002 | 0.7287 | 0.6825 | 0.7550 |
| **standard deviation** | 0.1668 | 0.1766 | 0.1692 | 0.1667 | 0.1394 |

According to the data of Table 8, students of the KM3 course achieved the best result in the solution of the exam test. The mean of this data series is the highest, while the standard deviation of it is the smallest. Considering means, the result was the worst at data series including results of solution of all tests of ABK1 course.

We decided with homogeneity tests related to means and standard deviations if the differences between means and standard deviations are significant. Our null hypotheses are:

*H0: The means are the same.*

*H0: The standard deviations are the same.*

We reject null hypotheses if the probability results of the tests are smaller than the significance level, otherwise we keep them. The data series do not have the same distribution if either the mean or standard deviation, or both of them are different. The result of that data series is significantly better, which has greater mean.

Table 9 shows the results of the homogeneity tests. According to the table, the mean of ABK1all and ABK2 all and mean of the ABK2 and the KM3 exam tests differ significantly. Although the standard deviations of these data series differ remarkably, we could not prove significant difference. In case of the other data series there is no significant difference between means and standard deviations.

Table 9. *Result of homogeneity tests*

| | **ABK1 all<br>ABK2 all** | **ABK1 exam<br>ABK2 exam** | **ABK1 exam<br>KM3 exam** | **ABK2 exam<br>KM3 exam** |
| --- | --- | --- | --- | --- |
| **standard deviation p/test/assessment** | 0.2366<br>Levene test<br>same | 0.8116<br>Levene test<br>same | 0.1498<br>Levene test<br>same | 0.3648<br>Levene test<br>same |
| **mean p/test/assesment** | 1.516e-06<br>Mann–Whitney test<br>not same | 0.173<br>Mann–Whitney test<br>same | 0.2846<br>Mann–Whitney test<br>same | 0.02343<br>Mann–Whitney test<br>not same |

We experienced that students of the KM3 course who learned from exercise book and achieved better result learned in groups and they also made notes in the course of consultations. This was not typical for the students who solved practise tests. Students of this course were more motivated then the students of the ABK courses. We asked them which method they prefer. By their answers the proportion is near same who prefer one or the other preparation manner [23].

We made the data mining examinations with open source AlphaMiner program. With the scores achieved by task and test trials of the ABK1 all data series, we made a decision tree by topic and by task type, to determine what scores are expected. See Table 10 and 11.

Table 10. Decision tree by task types, ABK1 all

| Task | Score | Confidence (%) | Support (%) |
|---|---|---|---|
| concept collation | 1 | 28.7 | 4.8 |
| giving definitions | 1 | 68.7 | 19.1 |
| knowledge of concept characteristics | 1 | 44.6 | 31.1 |
| concept identification | 1 | 47.8 | 9.3 |
| concept realization | 0 | 40.6 | 27.1 |
| mapping of the different model's concepts | 1 | 33.6 | 8.6 |

Table 10 shows that the least efficient step of concept learning is concept realization. In case of other task type the expected score is 1 point.

Table 11. Decision tree by topics, ABK1 all

| Topic | Score | Confidence (%) | Support (%) |
|---|---|---|---|
| E-R model | 1 | 32.0 | 17.8 |
| Dependences | 1 | 23.7 | 4.7 |
| Basic data modelling concept | 1 | 47.0 | 28.2 |
| Normalization | 0 | 51.1 | 26.8 |
| Relational model | 1 | 64.9 | 22.5 |

According to Table 11 you can expect the worst result at normalization, in the other topics the probable score is 1 point. For the ABK2 course the program could not make similar decision tree. It means that at every task type and topic the expected scores are the same. Because in the tests solutions most of the scores are 1 point (see frequency diagram of the Figure 5), probably the expectable score is 1 point in every task type and topic.

From the exam result of the KM3 course, we could make a decision tree by the number of tasks. This enables us to determine the expected scores by topics because this exam test includes task in the order of taught topics, see Table 12.

Table 12. Decision tree by number of the task, KM3 exam

| Rule | Score | Confidence (%) | Support (%) |
|---|---|---|---|
| task number: <= 14, topics: basic data modelling concepts, relational model, E-R model | 1 | 49.32 | 63.6 |
| task number: >14 and <=16, topics: dependences, dependency diagram, | 0 | 50.83 | 9.1 |
| task number: >17 and <=19, topics: normalization - primary key determination, how many normal forms are violated by dependences, in how many normal form is the table | 0 | 40.53 | 13.6 |
| task number: <19 ,topics: normalization - table decomposition, designation of the primary and foreign key | 1 | 44.26 | 13.6 |

Although the values of the confidences are not high, Table 12 confirms the opinion of the students that the most difficult topics are the dependences and the normalizations.

The results from the decision trees provided a good motivation to further analyze the scores achieved in trial tests. What could be the reason that most of the expected scores fall into the su-preme interval, and the worst expected scores fall into the bottom interval? We prepared a frequency diagram by the scores of exam test of a group of the KM3 course, see Figure 5.

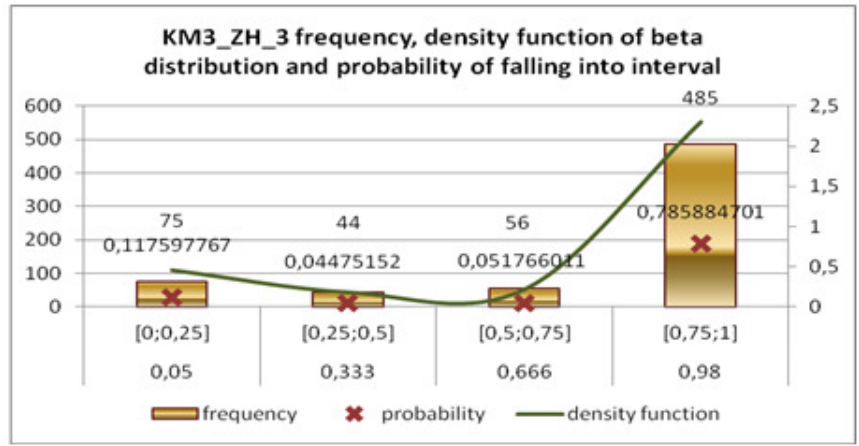

Figure 5. Distribution of scores achieved by questions and test trials

Figure 5 shows that 1 is the most frequent score the second most frequent is zero. The appearance of the frequency diagram and the fact that the scores are between 0 and 1 hints that the data has beta distribution. With Kolmogorov-Smirnov test we made a matching examination for homogeneity tests [24].

In the analyses undertaken it was observed that at zero value the difference is big, which is caused by the characteristics of the beta distribution function (at zero its value is also zero), but the calculated values differ from zero because there are a lot of results with zero points. We substituted the zero values with a small non-zero value of 0.05. It can be considered zero and it overcomes the experienced big difference. So we succeed to prove that scores by question of homogeneity test have generally beta distribution [25].

# 7. CONCLUSIONS

How effective is the concept teaching? Before the examinations we stated (see introduction), that we regarded those areas sufficient where the students achieved satisfactory or better result. According to the means of the exam tests, the result of all three courses reached this level. The result of the KM3 course almost measures up to the good result, it is only lacking a few tenth. Based on this we can say that the concept teaching is effective.

Among the topics, the teaching of dependences and normalization are most problematic. Students achieved the best result in basic modelling concepts and dependence diagrams topics. By task types concept realization and the mapping of the different model's concepts have the worst results, while students achieved the best result in knowledge of definitions and concept identification.

With the help of decision trees we could determine what result can be expected in certain topics and task types. At the problematic areas listed above, the expected score is 0, while at other areas it is 1. The reason behind the many 1 score results is the many 1 point scores for individual tasks.

After increasing the number of tasks in the Moodle question bank and in the tests, the results of all tests of the ABK2 course did become significantly better than the result of all test of the ABK1 course. We could not state the same for the results of the exam tests of the courses.

The mean of the result of the exam tests of those students who prepared for the exam with the help of printed exercise book is remarkably better. Its standard deviation is lower than the similar data of the results of the exam tests of students who prepared with electronic tests. The difference of the means is significant.

In our analysis we found that the scores of homogenous tests generally have beta distribution. This allows us to determine - based on estimated distribution - that in case of a certain test, what results and with what probability will students achieve.

It might worth to spend further research efforts to examine the knowledge of students about the connections between concepts, their ability to classify the concepts and to develop the hierarchical systems of the concepts. We already examined these both. In 2010, we executed a concept systematization survey where students had to define the connection between concepts by filling in empty hierarchical figures and Venn diagrams. Students filled in the hierarchical figures generally well, but we cannot say the same about Venn diagrams. In the hope of better results, we developed the concept maps of data modelling. In 2013, in a concept systematization survey, we examined if the concept systematization is significantly better with the use of concept maps. Students had to fill in the empty concept maps of the worksheet, and give the connection between concepts. The result was significantly better than the result of 2010 survey [13].

Furthermore, it can also be examined how students are able to use the concepts for describing situations and solving problems. It might be an interesting question, if the more effective concept learning leads to increase in the effectiveness of database design (E-R modelling and normalization). In this topic we did not carry out any examination yet. We examined the efficiency of the normalization, primarily focusing on alternative normalization methods that can help students normalize better than with conventional method [26].

Before the Moodle system and data modelling exercise book were available as educational tools, the concept learning meant learning definitions, looking for some examples, perhaps filling in tests on paper for students. The latter one can only be used once and its reproduction is difficult, its correction requires the teacher's contribution and the number of tasks is few. On the contrary, the tasks of Moodle only have to be developed once, they can be solved several times and their correction is automatic. You can also reuse the exercise book as an alternative preparation tool many times, if students write their answers on paper. The correctness of the results can also be checked by the students themselves based on the solution key given in the exercise book. Both tools support the steps of the concept learning listed above. We can recommend the usage of similar educational tools and concept systematization worksheets for every teacher. Based on our examinations, these tools make the teaching of concepts more effective.

## AUTHORS

Márta Czenky is an assistant professor in the Department of Informatics of Faculty of Mechanical Engineering of Szent István University (Gödöllő, Hungary). She is teaching informatics, programming, database management, database programming in BSC and MSC qualifications at more faculties. Her research area is teaching of database management, her publications deal mainly with this topic. More books of her were published in Hungary, among others: *Data modelling, application of SQL and Access, Let's learn informatics together, Programming in Access, Exercise book of data modelling tasks, Exercise book of Excel tasks.*

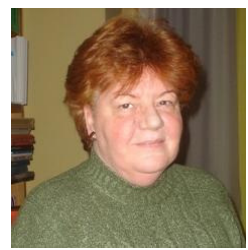